\documentstyle[aps,twocolumn,epsfig]{revtex}

\begin{document}
\draft

\twocolumn[\hsize\textwidth\columnwidth\hsize\csname@twocolumnfalse\endcsname

\title{Threshold temperature for pairwise and many-particle thermal entanglement in the isotropic Heisenberg model}
\author{Xiaoguang Wang}
\address{1. Institute for Scientific Interchange (ISI) Foundation, Viale Settimio Severo 65, I-10133 Torino, Italy, and}
\address{2. Department of Physics and Centre for Advanced Computing-Algorithms and Cryptography, \\
Macquarie University, Sydney, New South Wales 2109, Australia.}

\date{\today}
\maketitle

\begin{abstract}
We study the threshold temperature for pairwise thermal entanglement in the spin-1/2 isotropic Heisenberg model up to 11 spins and find that the threshold temperature for odd and even number of qubits approaches the thermal dynamical limit from below and above, respectively. The threshold temperature in the thermodynamical limit is estimated. We investigate the many-particle entanglement in both ground states and thermal states of the system, and find that the thermal state in the four-qubit model is four-particle entangled  before a threshold temperature.
\end{abstract}
]

Entanglement is a nonlocal quantum correlation and entangled states
constitute indeed a valuable resource in quantum information processing \cite
{Bennett}. Entanglement in systems of interacting spins \cite
{Connor01,Meyer01,Arnesen01,Wang01,Three} as well as in systems of
indistinguishable particles \cite{Schli,Li,Paolo,Paolo1,You} has been
investigated. In particular entanglement in both the ground state \cite
{Connor01,Meyer01} and thermal state \cite{Arnesen01,Wang01,Three} of a
spin-1/2 Heisenberg spin chain have been analyzed in the literature. The
intriguing issue of the relation between entanglement and quantum phase
transition \cite{QPT} have been addressed in a few quite recent papers \cite
{Osborne,Osterloh}.

The state we will consider is the thermal equilibrium state in
a canonical ensemble. The system state at finite temperature is given by the
Gibb's density operator $\rho _T=\exp \left( -H/kT\right) /Z,$ where $Z=$tr$%
\left[ \exp \left( -H/kT\right) \right] $ is the partition function, $H$ the
system Hamiltonian, $k$ is Boltzmann's constant which we henceforth will
take equal to 1, and $T$ the temperature. As $\rho_T$ represents a thermal
state, the entanglement in the state is called {\em thermal entanglement}%
\cite{Arnesen01}. It is important to stress that, although the central
object of statistical physics, the partition function, is determined by the
eigenvalues of $H$ only, thermal entanglement properties generally require
in addition the knowledge of the energy eigenstates. 
In some models such as the isotropic Heisenberg qubit rings the pairwise thermal entanglement can be completely determined by the partition function \cite{Victory}.  

Let us briefly review the main results of Ref. \cite{Victory}. We consider a
physical model of a ring of $N$ qubits interacting via the isotropic
Heisenberg Hamiltonian 
\begin{equation}
H=J\sum_{i=1}^{N-1}S_{i,i+1}+JS_{N,1},  \label{eq:xxx}
\end{equation}
where $S_{j,j+1}=${}$\frac 12\left( 1+\vec{\sigma}_i\cdot \vec{\sigma}%
_{i+1}\right) $ is the swap operator between qubit $i$ and $j$, $\vec{\sigma}%
_i=(\sigma _{ix},\sigma _{iy},\sigma _{iz})$ is the vector of Pauli
matrices, and $J$ is the exchange constant. In Ref. \cite{Victory} we proved
that there is no thermal entanglement between two nearest-neighbour qubits in the ferromagnetic ($J<0$) isotropic Heisenberg rings at any temperature. So we only
consider the antiferromagnetic case and the exchange constant $J$ is then
assumed to be 1. 
A direct relation is established between the concurrence\cite{Con} quantifying  two-qubit entanglement and a
macroscopic thermodynamical function, the internal energy $U$\cite{Victory}: 
\begin{equation}
C(N)=\max \left( 0,-\frac UN\right) =\max \left( 0,\frac 1{NZ}\frac{\partial
Z}{\partial \beta }\right),  \label{eq:cxxx}
\end{equation}
where $\beta=1/T$ and $C(N)$ refers to the concurrence for two nearest-neighbor qubits of the  $N$-qubit ring.  The
entanglement is uniquely determined by the partition function of the system.
Note that there is a slight difference between the expression of the
concurrence and the one given in Ref. \cite{Victory} due to a constant term
added in Eq.(\ref{eq:xxx}) for convenience of numerical calculations.
The reason why the entanglement is determined only by the eigenvalues of the system is that there are many symmetries in the isotropic Heisenberg model such as translational invariance and rotational symmetry along $x,y,z$ directions, et al. If we apply a nonzero magnetic field along $z$ direction we only have rotational symmetry along $z$ direction. In this case we can not determine the entanglement by eigenvalues alone.

From Eq.(\ref{eq:cxxx}) the concurrence for the ground state (GS) with
energy $E_{GS}$ is then followed by 
\begin{equation}
C_{GS}(N)=\max \left( 0,-\frac{{E_{GS}}}N\right) ,
\end{equation}
which is obtained in Ref. \cite{Connor01} for even-number qubits. But now we
know that it is valid for any number of qubits. From the available results 
\cite{Orba59,Bonner,Bar,Fab} of the ground-state energy we can determine the
concurrence directly. For the case of even-number qubits the concurrence is
given in Ref. \cite{Connor01}. Here in Table I we give the concurrence of
ground states for $N$ qubits up to $N=11$. {}From the table we find that the
entanglement increases with the increase of odd number $N$. On the contrary
the entanglement decreases with the increase of even $N$. We
also see that the ground state is always entangled except the case of $N=3$.

Since $\partial U/\partial T\geq 0$ and $C_{GS}(3)=0$ there is no
entanglement at any temperature for the case of $N=3$. For other cases, $C_{GS}(N)>0$ $ (N\neq 3),$ and the concurrence (\ref{eq:cxxx}) is a monotonic decreasing function of the temperature. It is then obvious that there exists a $N$-dependent threshold temperature $T_{th}(N)$ after which the
thermal entanglement disappears. The threshold temperature is determined by the equation $U(T_{th})=0$. Therefore the problem left is to examine the
threshold temperature and we  address it in this brief report. We will
determine the threshold temperature both analytically and numerically.

The number of qubits we will consider is up to $N=11$. Analytical results can be obtained for very small rings. For $N=2$, the eigenvalues of $H$ is easily found to be -2(1) and 2 (3)\cite
{Arnesen01}, where the number in the parenthesis is the degeneracy. For $N=3$
the eigenvalues of $H$ are 0(4) and 3(4)\cite{Three}. For $N=4,$ by direct
diagonalization, the eigenvalues are given by -2(1), 0 (3), 2(7) and 4(5) (see later discussions). From the eigenvalues the partition functions $Z(N)$ $(N=2,3,4)$ simply
follow as 
\begin{eqnarray}
Z(2) &=&e^{2\beta }+3e^{-2\beta },  \nonumber \\
Z(3) &=&4+4e^{-3\beta },  \nonumber \\
Z(4) &=&3+e^{2\beta }+7e^{-2\beta }+5e^{-4\beta }.  \label{eq:zzz}
\end{eqnarray}
The substitution of Eq.(\ref{eq:zzz}) into (\ref{eq:cxxx}) leads to 
\begin{eqnarray}
C(2) &=&\frac 1{Z(2)}\max \left( 0,e^{2\beta }-3e^{-2\beta }\right) , 
\nonumber \\
C(3) &=&\frac 1{Z(3)}\max \left( 0,-4e^{-3\beta }\right) =0,  \nonumber \\
C(4) &=&\frac 1{2Z(4)}\max \left( 0,e^{2\beta }-7e^{-2\beta }-10e^{-4\beta
}\right) ,  \label{eq:ccc}
\end{eqnarray}
which are explicit analytical expressions for the concurrence. The threshold
temperatures are directly obtained from the above equation: $T_{th}(2)=4/\ln
3\approx 3.641,$ $T_{th}(3)=0,$ and $T_{th}(4)=1.726728$.

For $N\geq 5$ we use numerical method to determine the threshold
temperature. We mention that some analytical results of the eigenvalue problem
for $N=5,6$ and $7$ can be found in Ref. \cite{H567}. 
Bethe's Ansatz\cite{Bethe} can be used for the calculations of eigenvalues. Here we use the exact  diagonalization method to calculate the eigenvalues. The obvious symmetry $[H,S_z]=0$ allows to decompose the Hilbert space of the system into a set of orthogonal subspaces. Here $S_z=1/2\sum_{i=1}^N\sigma_{iz}$ is the collective operator for $N$ qubits.
The dimension of the subspace is much smaller than the original Hilbert space, and then we can diagonalize the Hamiltonian in the subspaces.

The numerical results for the threshold temperature are given in Table II and III, from which we
find that threshold temperature increases (decreases) as $N$ increases for
the case of odd (even) qubits. We also observe that the threshold
temperature converges very quickly. Then we can estimate the threshold
temperature in the thermodynamic limit $T_{th}(\infty )\approx \allowbreak 1.\,5903$ from the
numerical results of these {\it finite systems}. So we have the following relation
\begin{equation}
T_{th}(N)|_{N\text{ odd}}<T_{th}(\infty )<T_{th}(N)|_{N\text{ even}},
\end{equation}
which implies that the threshold temperatures for $N\,$odd and $N$ even
approach the thermal dynamical limit from below and above, respectively.

We have considered only pairwise thermal entanglement of a multiqubit
system. Another type of entanglement is the $N$-particle 
entanglement which involves all $N$ particles. As far as we know there are no discussions on many-particle thermal entanglement.
In the following we will consider the many-particle entanglement in the ground states of the Heisenberg models up to 11 qubits and then study the many-particle thermal entanglement within the four-qubit Heisenberg model.

For a three-qubit pure state, there exists a good measure of three-particle entanglement\cite{Cof00}. Here we use the {\em state preparation fidelity} $F$ of a $N$-qubit state $\rho$ to investigate the many-particle entanglement. The state $\rho$ can be either pure or mixed. The fidelity is defined as\cite{Sackett}
\begin{equation}
F(\rho)=\langle\Psi_{\text{GHZ}}| \rho|\Psi_{\text{GHZ}}\rangle,
\label{eq:ff}
\end{equation}
where $|\Psi_{\text{GHZ}}\rangle=1/\sqrt{2}(|00...0\rangle+|11...1\rangle)$ is the $N$-particle Greenberger-Horne-Zeilinger (GHZ) state\cite{GHZ}.

The sufficient condition for $N$-particle entanglement is given by\cite{Sackett}
\begin{equation}
{F} (\rho)>1/2.
\end{equation}
We have a freedom to choose other GHZ states such as 
$|\Psi_{\text{GHZ}}\rangle=1/\sqrt{2}(|00...01\rangle\pm |11...10\rangle)$, etc.
By local unitary operations we can transfer these states to the original GHZ state and the operations do not change the entanglement. Detail discussions on this sufficient condition can be found in Ref.\cite{Sackett,Uffink}. 

Let us study the many-particle entanglement in the ground states of the Heisenberg models for the number of qubits up to 11. It is known that \cite{Bar} the ground state which has $S_z=0$, is non-degenerate if $N$ is even, and the ground state which has $S_z=\pm 1/2$, is four-fold degenerate if $N$ is odd. 
Then the ground state is a pure (mixed) state for even (odd) $N$. We numerically find the ground state of the system, from which we calculate the fidelity. The form of the GHZ state is choose in order that the fidelity is as large as possible. The numerical results is given in Table IV. First we observe that the fidelity of four-qubit ground state is larger than one half, hence it is a many-particle entangled state. The fidelity is less than one half for $N$ from 2 to 11 except $N=4$. If we consider even or odd number $N$, the fidelity decreases as $N$ increases. From the analytical expressions of the ground states for $N=3$\cite{Three} and $N=4$ (see below), we can analytically obtain the fidelity $F=1/6$ for $N=3$ and $F=2/3$ for $N=4$, respectively.

Having  found that the ground state of the four-qubit Heisenberg model is a 4-particle entangled state, we expect that the thermal state will be also a 4-particle entangled state for small temperature. We also expect a threshold temperature before which the ground state is 4-particle entangled. 
To study the many-particle thermal entanglement we need to know all the eigenvalues and eigenstates. For convenience following notations are used
\begin{eqnarray}
|n\rangle _0&=&{\cal T}^{n-1}|1\rangle _0 ,|1\rangle
_0=|1000\rangle, \nonumber\\ 
|n\rangle _1&=&{\cal T}^{n-1}|1\rangle _1(n=1,2,3,4), |1\rangle _1 =|1100\rangle ,\nonumber \\
|n\rangle _2&=&{\cal T}^{n-1}|1\rangle _2 (n=1,2), |1\rangle _2 =|1010\rangle.
\end{eqnarray}
Here ${\cal T}$ is a unitary cyclic right shift operator\cite{Bar} defined by its action on the basis ${\cal T}|m_1m_2m_3m_4\rangle=|m_4m_1m_2m_3\rangle$ and 
$|m_1m_2m_3m_4\rangle=|m_1\rangle\otimes|m_2\rangle\otimes|m_3\rangle\otimes|m_4\rangle$.

A direct diagonalization of the four-qubit Hamiltonian gives all    eigenvectors and eigenvalues,
\begin{eqnarray}
E_0&=&4, |\Psi_{0}\rangle= |0000\rangle, \nonumber\\
E_{15}&=&4, |\Psi_{15}\rangle= |1111\rangle,  \nonumber\\
E_k&=&(2+t_k+t_k^{-1}),|\Psi_k\rangle=\frac 1{2}
\sum_{n=1}^4t_k^{-n}|n\rangle _0, \nonumber\\
E_{k+4}&=&(2+t_k+t_k^{-1}),|\Psi_{k+4}\rangle=\frac 1{2}
\sum_{n=1}^4t_k^{-n}\Lambda_x|n\rangle _0, \nonumber\\
E_{9} &=&-2,|\Psi_{9}\rangle =\frac 1{2\sqrt{3}}\left(
\sum_{n=1}^4{\cal T}^{n-1}|1\rangle _1-2\sum_{n=1}^2{\cal T}^{n-1}|1\rangle _2\right), 
\nonumber \\
E_{10} &=&0,|\Psi_{10}\rangle =\frac 1{\sqrt{2}}(|1010\rangle -|0101\rangle ), 
\nonumber \\
E_{11} &=&2,|\Psi_{11}\rangle =\frac 1{\sqrt{2}}(|1100\rangle -|0011\rangle ), 
\nonumber \\
E_{12} &=&2,|\Psi_{12}\rangle =\frac 1{\sqrt{2}}(|1001\rangle -|0110\rangle ), 
\nonumber \\
E_{13} &=&2,|\Psi_{13}\rangle =\frac 12(|1100\rangle +|0011\rangle -|1001\rangle
-|0110\rangle ),  \nonumber \\
E_{14} &=&4,|\Psi_{14}\rangle =\frac 1{\sqrt{6}}\left(
\sum_{n=1}^4{\cal T}^{n-1}|1\rangle _1+\sum_{n=1}^2{\cal T}^{n-1}|1\rangle _2\right),
\label{eq:state2}
\end{eqnarray}
where the parameters $t_k=\exp (i\pi k/2) (k=1,2,3,4)$, and the operator
$\Lambda_x=\sigma_x^{\otimes  4}=\sigma_x\otimes\sigma_x\otimes\sigma_x\otimes\sigma_x$.

These eigenvalues and eigenstates completely determine the thermal state $\rho_T$. We let our GHZ state be $|\Psi_{\text{GHZ}}\rangle={1}/{\sqrt{2}}(|1010\rangle+|0101\rangle)$.
Then from Eq.(\ref{eq:ff}) and all the eigenvalues and eigenstates we get
\begin{equation}
F(\rho_T)=\frac{e^{-4\beta}/3+2e^{2\beta}/3}{3+e^{2\beta}+7e^{-2\beta}+5e^{-4\beta}}. \label{eq:fffff}
\end{equation}
It is easy to verify that $F=2/3$ in the limit of $T\rightarrow 0$ as we expected.

In Fig.1 we give the numerical result for the state preparation fidelity against temperature. For comparison we also plot the concurrence. First we observe that 
there exists a threshold temperature $T_{th}$ for the state preparation fidelity before which the thermal state $\rho_{T}$ is four-particle entangled state. The threshold temperature $T_{th}\approx 0.83$. Since the condition $F>1/2$ is a sufficient condition we can not say that the thermal state has no four-particle entanglement after $T_{th}$. From the figure we can see that the thermal state has both pairwise entanglement and four-particle entanglement before the threshold temperature. But both entanglement are not maximal. 

In conclusion we investigated the threshold temperature for the pairwise thermal
entanglement in the isotropic Heisenberg model with finite number of qubits
up to $N=11$. We find that the threshold temperatures for odd and even
number of qubits approach the thermal dynamical limit from below and above, respectively.
Although we only make calculations of the threshold temperatures for small
spin rings we can well estimate the threshold temperature in the
thermodynamical limit as $T_{th}(\infty )\approx \allowbreak 1.\,5903$ since 
$T_{th}(N)$ converges very quickly as $N$ increases. 
By using one sufficient condition for many-particle entanglement, the state preparation fidelity is larger than one half, 
we investigated the many-particle entanglement of both ground states and thermal states in the Heisenberg models. We find that the ground state of the four-qubit model is a four-particle entangled state, and also find that there exists a threshold temperature $T_{th}\approx 0.83$ before which the thermal state is a four-particle entangled state. It is interesting to see that pairwise entanglement and four-particle entanglement (not maximal) coexists in this system.

{\em Acknowledgments}
The authors thanks for the helpful discussions with Paolo Zanardi, Irene D'Amico, Barry C Sanders, and Prof. V.E.Korepin, H. Fu, A. I. Solomon, and Guenter Mahler. This work has been supported by the European Community through grant IST-1999-10596 (Q-ACTA).

\begin{table}[tbp]
\caption{The ground-state concurrence for the number of qubits from 2 to 11.}
\begin{tabular}{ccccccccccc}
$N$   & 2  & 3   & 4    & 5    & 6   & 7     &8       & 9     &10    & 11         \\ \hline
$C_{GS}$ & 1.0& 0.0 & 0.5  & 0.247 & 0.434 & 0.316 &0.412  & 0.344&0.403 & 0.358 
\end{tabular}
\end{table}

\begin{table}[tbp]
\caption{The threshold temperature for even-number qubits}
\begin{tabular}{cccccc}
$N$ & 2 & 4 & 6 & 8 & 10 \\ \hline
$T_{th}$ & 3.64095691 & 1.72672823 & 1.60976354 & 1.59167655 & 1.59038369
\end{tabular}
\end{table}

\begin{table}[tbp]
\caption{The threshold temperature for odd-number qubits}
\begin{tabular}{cccccc}
$N$ & 3 & 5 & 7 & 9 & 11 \\ \hline
$T_{th}$ & 0.0 & 1.53825517 & 1.58556286 & 1.58979598 & 1.59020107
\end{tabular}
\end{table}

\begin{table}[tbp]
\caption{The state preparation fidelity of the ground states for the number of qubits from 2 to 11.}
\begin{tabular}{cccccc}
$N$  & Fidelity & GHZ states \\ \hline
$2$  & 1.0000 & $|01\rangle-|10\rangle$  \\ \hline 
$3$  & 0.1667 & $|010\rangle+|101\rangle$ \\ \hline
$4$  & 0.6667 & $|0101\rangle+|1010\rangle$  \\ \hline 
$5$  & 0.0873 & $|01010\rangle+|10101\rangle$ \\ \hline 
$6$  & 0.4580 & $|010101\rangle-|101010\rangle$ \\ \hline 
$7$  & 0.0505 & $|0101010\rangle+|1010101\rangle$ \\ \hline 
$8$  & 0.3173 & $|01010101\rangle+|10101010\rangle$ \\ \hline 
$9$  & 0.0306 & $|010101010\rangle+|101010101\rangle$ \\ \hline 
$10$ & 0.2205 & $|0101010101\rangle-|1010101010\rangle$\\ \hline 
$11$ & 0.0191 & $|01010101010\rangle+|10101010101\rangle$ 
\end{tabular}
\end{table}

\begin{figure}
\begin{center}
\epsfig{width=10cm,file=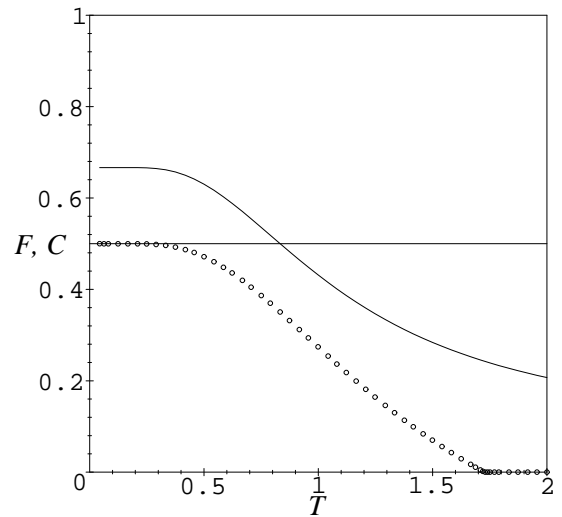}
\caption[]{The state preparation of fidelity (solid line) and the concurrence (circle dots) against the temperature.} 
\end{center}
\end{figure}

\end{document}